\documentstyle[12pt,epsfig]{article}
\newcommand{\be}{ \begin{eqnarray}}
\newcommand{\ee}{\end{eqnarray}}
\newcommand{\beno}{ \begin{eqnarray*}}
\newcommand{ \eeno}{\end{eqnarray*}}
\newcommand{\bfg}{\begin{figure}}
\newcommand{\efg}{\end{figure}}

\begin{document}

\centerline{\bf\Large  Can  dileptons be observed in heavy ion collisions at RHIC? }
\vskip 1cm
\centerline{\bf E. Shuryak}
\centerline{\it Physics Department, State University of New York}
\centerline{\it Stony Brook, NY 11794}
\vskip 1cm

\begin{abstract}
Both dilepton and charm production at RHIC are considered to be 
important signatures for Quark-Gluon Plasma production. Recently it was
argued  by S.Gavin, P.L.McGaughey,P.V. Ruuskanen and
R.Vogt \cite{GMRV}
that the background from semileptonic correlated charm decays is so
large that it makes dilepton measurements virtually impossible. We
show that this conclusion in fact is  in fact
reversed if the energy loss due
to secondary interaction of charmed quarks is included.

\end{abstract}

   Dileptons produced in highly excited hadronic matter
provide valuable information about the hottest and the most
dense stages of nuclear collisions. Together with
photons, they  are the so called {\it penetrating
probes}  \cite{Shuryak_80} which suffer very little secondary
interaction.
Consequently, dilepton measurements 
   has attracted great deal of attention, both of theorists and
   experimentalists. Referring specifically to highest energies, let us
remind that one of
the major RHIC detectors, PHENIX, and ALICE at LHC plan dilepton
measurements, both with electron and muon
pairs.

   Another potential QGP signature is {\it thermal production of new quark
     flavors}, especially of charm \cite{Shuryak_80,Muller,Gyulassy}
\footnote{ The expected highest temperatures
     at RHIC  \cite{Shuryak_80,hotglue,Geiger} are $T_i=400-500
     MeV$, so the mean energy per parton $\approx 3 T$ is comparable
     to charm quark mass. Note also that 
the mass of the strange
     quark is not large enough to suppress its production in the
hadronic phase.} However, the ordinary partonic production of charm
at the first impact is large, and whether it dominates the secondary
(and  thermal) charm production remains unclear. Nevertheless,
  charm signal will also be experimentally addressed at RHIC, by
STAR and PHENIX collaborations. 

\begin{figure}[h]
\begin{center}
\leavevmode
\epsfig{file=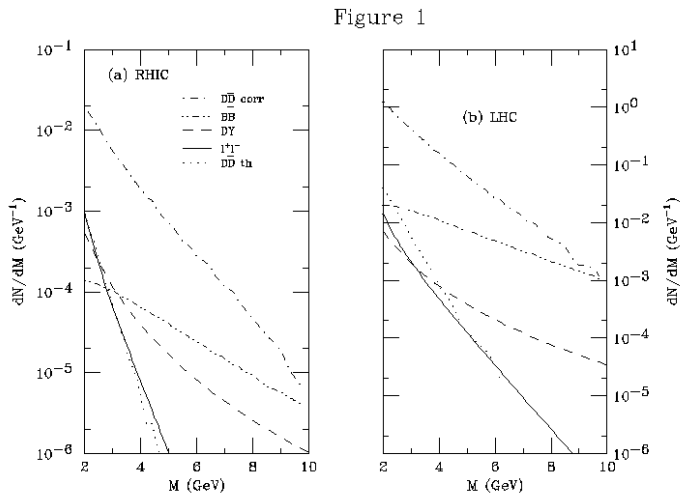, width=4in}
\end{center}\caption{\label{ramona} 
Contributions of different dilepton production mechanisms according to
GMRV  for central  Au Au collisions at RHIC (a) and LHC (b).
The curves  correspond to: a correlated
charm decays (dash-dots), b-quark decays
(dash-double dots), Drell-Yan process (dashed curve), thermal dileptons
(solid curve)
and  decays of thermally produced charm (dots). 
}
\end{figure}

   Since charmed  hadrons have substantial semileptonic decays,
their simultaneous decay create a $l^+l^-$
background for dilepton measurements.
This issue was addressed recently by S.Gavin, P.L.McGaughey,P.V. Ruuskanen and
R.Vogt
 (below GMRV) in a  detailed paper
   \cite{GMRV}. Their conclusions are summarized in Fig.\ref{ramona},
   and they
basically imply that the background
from leptonic decays of charmed (and even b) quarks is so large, that
implementation of the dilepton measurements is virtually impossible in
the whole kinematic domain.

  In this short paper we   question those pessimistic conclusions, and
  suggest
that
 it should be reversed. We show
  that a very
 important effect is missing from the GMRV analysis: unlike dileptons,
 the charmed quarks are not ``penetrating probes'', and
their spectra should be very
different
in pp and heavy ion collisions.  
 Like any other
 quarks (and gluons), charmed ones are also 
 subject to energy losses due to
multiple secondary interactions in dense matter produced in the collisions.
As we will show below, they are mostly stopped in matter.

  Our first point  is a purely geometrical observation. Consider
collision of two heavy nuclei, and imagine that in it a $\bar c c$
pair is produced\footnote{The commonly used terminology 
 a $correlated$ and an $uncorrelated$ charm decay. The former
is a simultaneous decay into $l^+$ and  $l^-$ from a  $\bar c c$
pair produced in one parton collisions,
while the latter comes from the charm quarks produced $independently$.
In this note we concentrate on the correlated background
only because the uncorrelated background can be statistically subtracted in
a standard way.
}.  The charmed quarks have to pass certain distances
$d_1, d_2$  on their way out, and we point out  that it is very
improbable that the 
 $sum$ $d_1+ d_2$ is small because the quarks are mostly
 produced back-to-back.

\begin{figure}[h]
\begin{center}
\leavevmode
\epsfig{file=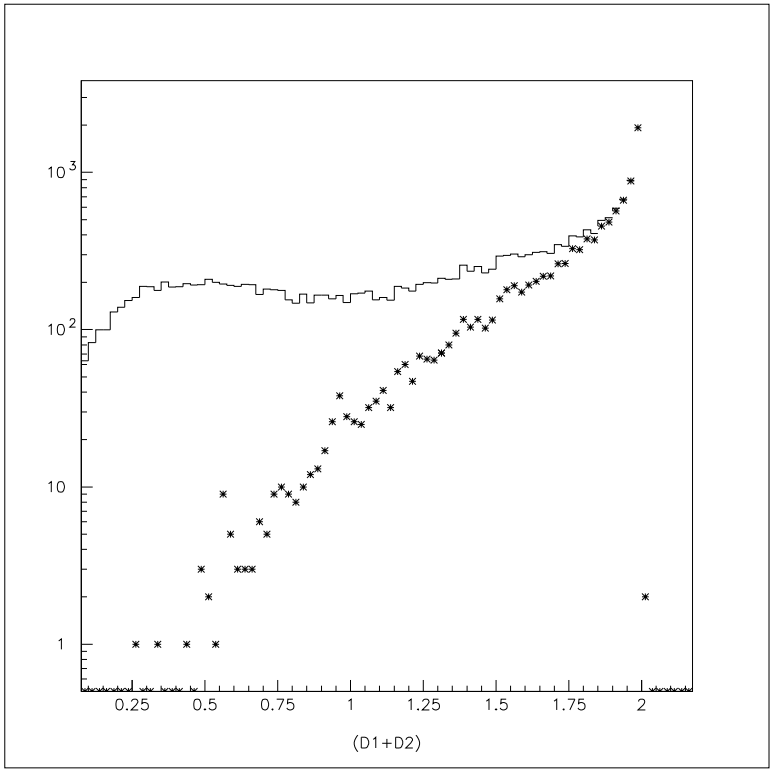, width=4in}
\end{center}\caption{\label{d1d2}
The histogram shows distribution of the
transverse distance d passed by a charm quark on the way out of nuclei
(in units of its radius), while stars correspond to the sum
$d_1+ d_2$ of distances for charmed quark and anti-quark.}
\end{figure}

If nuclei are approximated as spheres with a
well-defined surface and radius
R, one can easily quantify the relevant distributions. A distribution
over a single
quark path d (in units of the nuclei radius R) is shown by 
a histogram
in Fig.\ref{d1d2}. Note that it is basically flat between  d$\approx
.2 R$
(or 1 fm for heavy nuclei)
and $2R$ (the diameter). However, the 
 distribution of $(d_1+ d_2)/R$
(shown by stars in Fig.\ref{d1d2})
 is quite different. It is sharply
  peaked at its largest value, but is very strongly suppressed
at small ones. In order for both charmed quarks to escape, they 
not only  should be created
close to the surface, but also  quarks should be emitted
  in a very
small (tangent) solid angle. As we will show
shortly, 
this simple observation is in fact
responsible for a 
significant reduction of the correlated charm (and bottom)-induced background.

  A dynamical ingredient
of our analysis is $dE/dx$, the  quark energy losses in QGP. We would
not
comment here on a
 complicated history of its discussion in theoretical and
 phenomenological
papers. A 
 consistent treatment (generalizing
Landau-Pomeranchuck-Migdal approach to QCD) was recently developed
 by \cite{BDPS} (BDPS). The main qualitative difference between QED
 and QCD cases can  briefly be explained as follows.
In QED an electron is scattered and has a complicated zigzag-like trajectory, 
while its field go without interaction by a straight line.
In QCD it is the quark which is
going  by approximately straight line, while its gluonic field suffer
multiple rescatterings. 
The BDPS result for the energy loss is
\be
{dE\over dx}=C_R \alpha_s ({E\mu^2 \over \lambda_g})^{1/2} log( {E
\over \lambda_g \mu^2}) 
\ee
 where $C_R$ is Casimir for quark color representation, E is the
collision energy, $\lambda_g$ is $gluonic$ mean free path 
and $\mu$ is the rms momentum transfered in each
scatterings. Substituting some ``reasonable'' parameters of QGP at
RHIC (corresponding to ``hot glue scenario, see \cite{hotglue,Geiger})
we have estimated  {$dE/dx\approx 2 GeV/fm$.
  
  Our next step is Monte Carlo simulation of charm production. In
  order not to introduce any additional points of discussion, we 
follow ref.\cite{GMRV} as close as possible. We
have ignored the ``thermal charm'' and assumed that each
central
AuAu collision at RHIC produces $\bar c c$ pairs with a
very stiff $p_t$ distribution\footnote{An approximate parameterization
  used
is $dN/dp_t^2 \sim
1/(p_t^2+0.5)^{2.2}$, so the power of $p_t$ is close to 4.}  
generated by the leading order and
resummed next-to-leading-order QCD processes.
This approach works in pp case, and that is why  GMRV has found that
the correlated charm decay contributes so strongly at large dilepton masses.
  
However,
  after the energy losses dE/dx are included, 
only very few of  c or b quarks can in fact 
escape, while most of them are stopped.
Eventually, those should have $p_t$ spectra similar to all other hadrons,
governed by low decoupling temperature $T\sim 140 MeV$ and hydro
effects. Since both thermal and hydro velocities are not large, we
have ignored them.

 We have simulated semileptonic decays of c and b quarks and show
the resulting invariant mass $M=(p_{l+}+p_{l-})^2$
spectrum in Figs.\ref{mc}, \ref{mb}. In both cases
the histogram shows free decays, while
stars  include the effect of dE/dx. Those two cases are very different:
 while in free space the invariant mass
distribution has a smooth and large tail toward the large masses, 
 with dE/dx one clearly see
two distinct components:  charm decay at
rest and the contribution of escaping ones. The  boundary 
between two components is   at
$M_{l^+l^-}\approx 1.7 GeV$ for c and
4.5 GeV for b decays. Above it we have found a background
suppression, roughly by about 2 orders of 
magnitude. These features survive
 reasonable modification of charm production
spectra or of the chosen dE/dx value.

\begin{figure}[h]
\begin{center}
\leavevmode
\epsfig{file=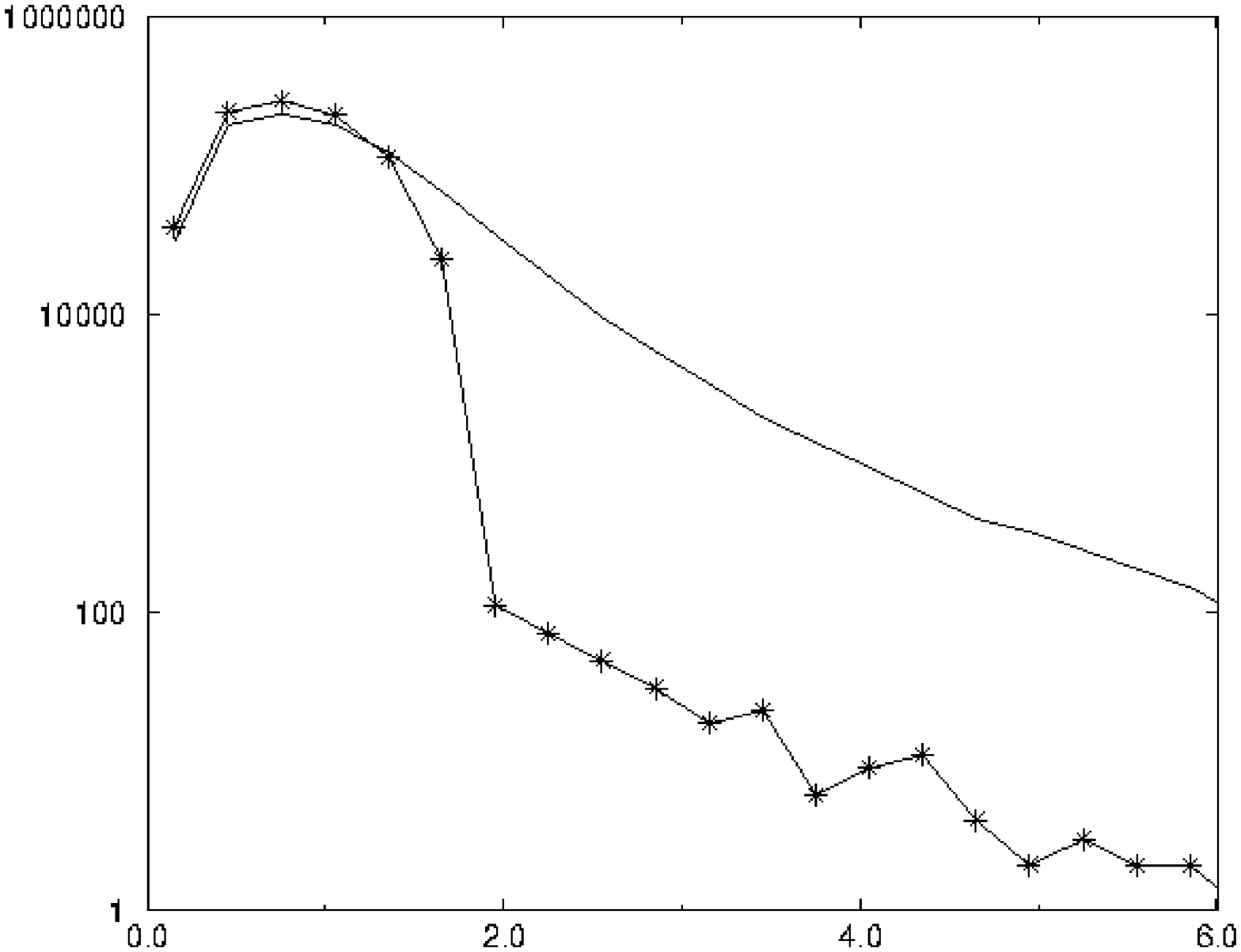, width=3.5in}
\end{center}\caption{\label{mc}
The distribution of dilepton invariant masses produced by a
semileptonic decays of charmed quarks, with (stars) and without
(histogram)
the matter effect due to dE/dx.
}
\end{figure}

\begin{figure}[h]
\begin{center}
\leavevmode
\epsfig{file=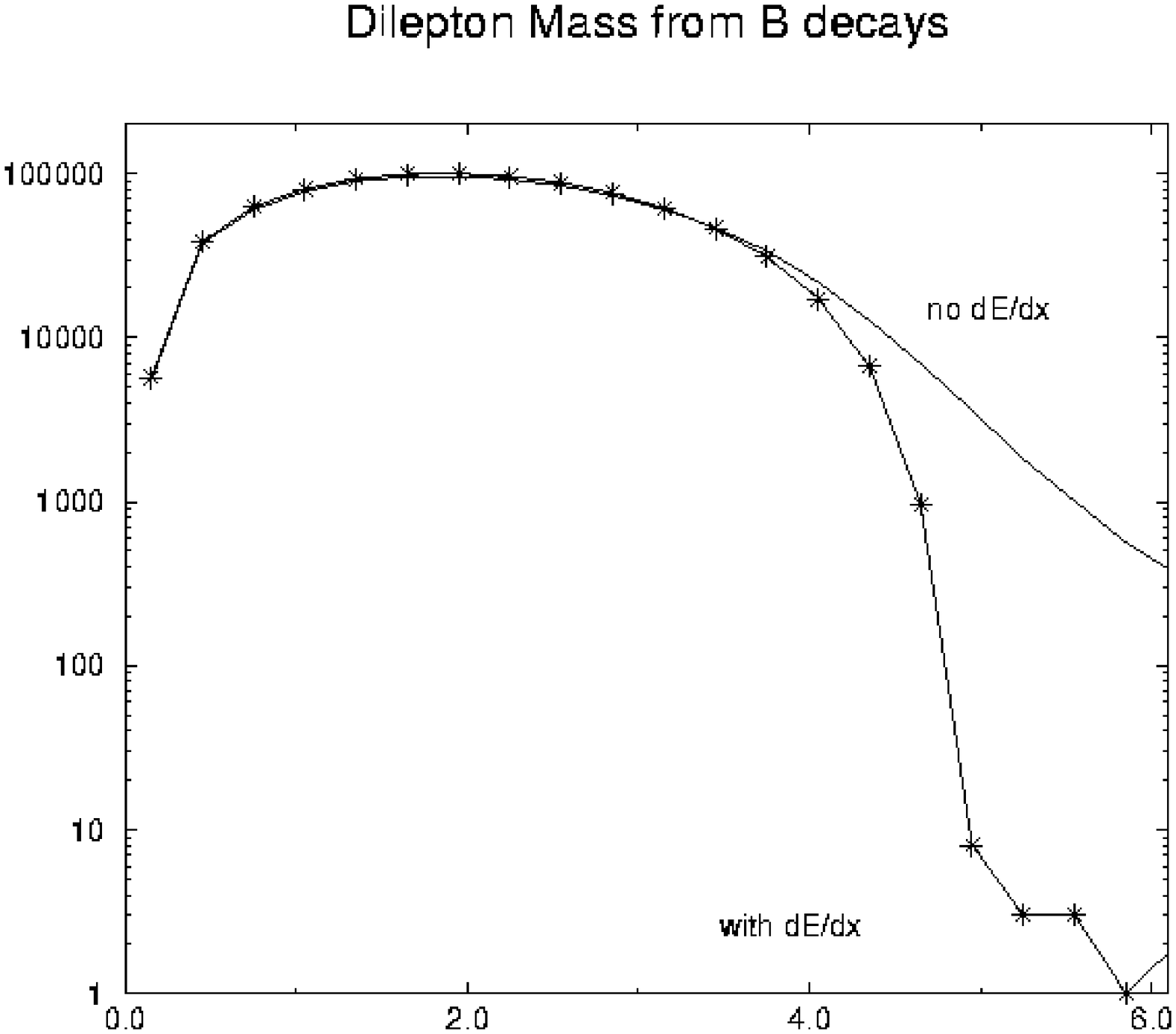, width=3.5in}
\end{center}\caption{\label{mb}
Same as the previous figure, but for b quark decays.
}
\end{figure}

  How important this reduction may be in practice? 
In order to answer this question, one has to evaluate dilepton
production,
both primary (known as the Drell-Yan process) and secondary
(non-equilibrium \cite{Geiger_Kapusta} and thermal \cite{Shuryak_Xiong}) ones. 
  In this short note we will not go into discussion of it\footnote{
Let us only mention that GMRV make a very good job on DY,  but 
do not include the non-equilibrium one. Also, they
treat  thermal in the leading order only. Both effects are expected to
increase the secondary production substantially.} 
and simply return to GMRV estimates. As seen 
from Fig.\ref{ramona} 
the ratio (dilepton yield)/(correlated charm
background) is about 1/10 for $M=2-8 GeV$, while  (dilepton yield)/(b decay
background) is about 1/3 for $M > 5 GeV$. Those are exactly the mass regions
where our
suppression  discussed above appears! Thus we  conclude that
(dilepton yield)/(correlated charm
background) is probably above 1; and that b-decays are
simply negligible. More quantitative conclusion is difficult to get now:
also one should consider acceptance of the particular detector, etc.
        
  Since we are still in a situation with the
   signal/background ratio being around 1, 
additional experimental tools are needed in order to separate dileptons from
charm decays. At least two are available: (i) 
dileptons are produced back-to-back in azimuthal angle,
 while leptons from charm decay are
nearly isotropic in it; (ii) Drell-Yan pairs have the well known
$(1+cos^2\theta)$ distribution where $\theta$ is the polar angle
between the dilepton direction in its CM frame and the beam. 
Also DY and direct charm should have simple scaling $A^{4/3}$ from
light nuclei (or peripheral collisions), so any excess over it is an
indications for secondary processes.

   In summary: in contrast to  GMRV, we think that c and b quarks
 produced in high energy heavy ion collisions  should
be  trapped in matter with very high probability. As a result, the
   background
due to correlated semileptonic charm decay does $not$ dominate the
dilepton spectra for invariant masses above 2 GeV.  
Optimistically, by using various
angular distributions, one may probably
measure
$both$ dileptons and charm.

\vskip 1cm
{\bf Acknowledgments} 
This work was stimulated by a seminar by Ramona Vogt: I am
thankful for her detailed
explanations of their work.
This work is partly supported by the 
US Department
 of Energy under Grant No. DE-FG02-88ER40388.

\end{document}